\documentclass[aps,pra,twocolumn,floatfix,superscriptaddress]{revtex4-1}
\usepackage{amsmath}
\usepackage{amssymb}
\usepackage{xcolor}
\usepackage{graphicx}
\usepackage{dcolumn}
\usepackage{braket}

\begin{document}

\title{Dynamical signatures of bound states in waveguide QED}

\author{E. S\'anchez-Burillo}
\affiliation{Instituto de Ciencia de Materiales de Arag\'on and Departamento de F\'isica de la Materia Condensada, CSIC-Universidad de Zaragoza, E-50009 Zaragoza, Spain}
\author{D. Zueco}
\affiliation{Instituto de Ciencia de Materiales de Arag\'on and Departamento de F\'isica de la Materia Condensada, CSIC-Universidad de Zaragoza, E-50009 Zaragoza, Spain}
\affiliation{Fundaci\'on ARAID, Paseo Mar\'ia Agust\'in 36, E-50004 Zaragoza, Spain}
\author{L. Mart\'in-Moreno}
\affiliation{Instituto de Ciencia de Materiales de Arag\'on and Departamento de F\'isica de la Materia Condensada, CSIC-Universidad de Zaragoza, E-50009 Zaragoza, Spain}
\author{J. J. Garc\'ia-Ripoll}
\affiliation{Instituto de Fisica Fundamental, IFF-CSIC, Calle Serrano 113b, E-28006 Madrid, Spain}
\begin{abstract}
We study the spontaneous decay of an impurity coupled to a linear array of bosonic cavities forming a single-band photonic waveguide. The average frequency of the emitted photon is different from the frequency for single-photon resonant scattering, which perfectly matches the bare frequency of the excited state of the impurity. We study how the energy of the excited state of the impurity influences the spatial profile of the emitted photon. The farther the energy is from the middle of the photonic band, the farther the wave packet is from the causal limit. In particular, if the energy lies in the middle of the band, the wave packet is localized around the causal limit. Besides, the occupation of the excited state of the impurity presents a rich dynamics: it shows an exponential decay up to intermediate times, this is followed by a power-law tail in the long-time regime, and it finally reaches an oscillatory stationary regime. Finally, we show that this phenomenology is robust under the presence of losses, both in the impurity and the cavities.
\end{abstract}

\maketitle

\section{Introduction}

Interactions between few-level systems (or quantum impurities) and photonic media with nonlinear dispersion relations and band gaps give rise to a plethora of interesting phenomena \cite{Lambropoulos2000}. Examples are the modification of the level structure of the impurity \cite{John1990,John1991,Camacho2015}, non-trivial dynamics \cite{Khalfin1958,Bykov1975,Fonda1978,Hack1982,Onley1992,John1994,Gaveau1995,Garmon2013,Redchenko2014,Lombardo2014}, and charge transfer enhancement \cite{Tanaka2006}. %
A characteristic phenomenon is the appearance of bound states\ \cite{Bykov1975,John1984,John1987} where a photonic excitation is confined to the vicinity of the impurity. This idea has been studied in various theoretical works, finding phenomena such as suppression of decoherence \cite{Tong2010a}, preservation of quantum correlations \cite{Tong2010b,Yang2013,Lu2013}, or the existence of multi-photon bound states\ \cite{Calajo2016,Calajo2016b,Shi2016}. An instance of bound state has been experimentally found \cite{Liu2017} in a circuit QED architecture \cite{Astafiev2010,Hoi2011,Hoi2013,VanLoo2013,Hoi2013b}, and effects of band gaps in qubit-qubit interactions have been measured \cite{Hood2016} in photonic crystals \cite{Arcari2014,Sollner2015,Lodahl2015}. There are other state-of-the-art technologies where these states can be potentially detected, \emph{e.g.} cold atoms \cite{thompson2013,goban2015} and diamond structures with color centers \cite{Sipahigil2016,Bhaskar2017}.

In this work, we study the signatures of bound states in the spontaneous decay in a waveguide-QED scenario. We choose a prototypical model where an impurity is coupled to a bosonic medium: a tight-binding model, which gives a cosine-shaped band. Due to the finite width of the band, two states appear bound to the impurity. This problem has already been treated in the literature when the energy of excited state of the impurity is in the middle of the band \cite{Lombardo2014} or when it is close to its inferior limit, so the superior limit of the band can be neglected \cite{Garmon2013}. Here, we solve it for \emph{general} values of the parameters, which are the energy of the excited state of the impurity with respect to the the band and the ratio between the impurity-photon coupling and the bandwidth.

We find an energy shift of the emitted photon with respect to the energy required to excite the impurity, provided the latter is not in the middle of the photonic band. Naively, one could argue that spectral features in the spontaneous emission should also appear in the scattering, since a scattering process comprises both absorption and emission of the photon by the impurity. However, there is no shift in the single-photon scattering \cite{Nori2008a, Fan2005a, Fan2005b}. 
Secondly, we study the spatial profile of the emitted photon and discuss the differences with respect to a photonic medium with a linear dispersion relation. Lastly, we find a rich dynamics in the excited state of the impurity. First, it decays exponentially with a decay rate different from that given by the Fermi's golden rule, oscillating with a phase which is shifted with respect to the bare energy of the impurity. This shift, which corresponds to the Lamb effect, is different from the shift found in the energy of the emitted photon. After the initial exponential decay, the dynamics presents an algebraic decay due to the presence of singularities in the density of photonic states. These power tails are robust under the presence of losses, both in the impurity and the cavities. Eventually, it reaches a stationary oscillating regime. 

The manuscript is organized as follows. In Sect. II, we introduce the Hamiltonian and summarize both its spectrum in the single-excitation subspace and its one-photon-scattering properties. In section \ref{sec:spontaneous_decay}, we discuss the main results of the paper. First, we present the already mentioned frequency shift of the emitted photon as a function of the coupling constant and the energy of the excited state of the impurity. Then, we study the spatial distribution of the emitted photon. We next characterize the spontaneous emission when the impurity is initially excited and discuss the effect of the losses.
We end up with the conclusions in Sect. IV. Some technical details are described in the appendices.

\section{Model}\label{sec:model}

\subsection{Hamiltonian and bound states}\label{sec:H_BS}

\begin{figure}[thb!]
\includegraphics[width=1\columnwidth]{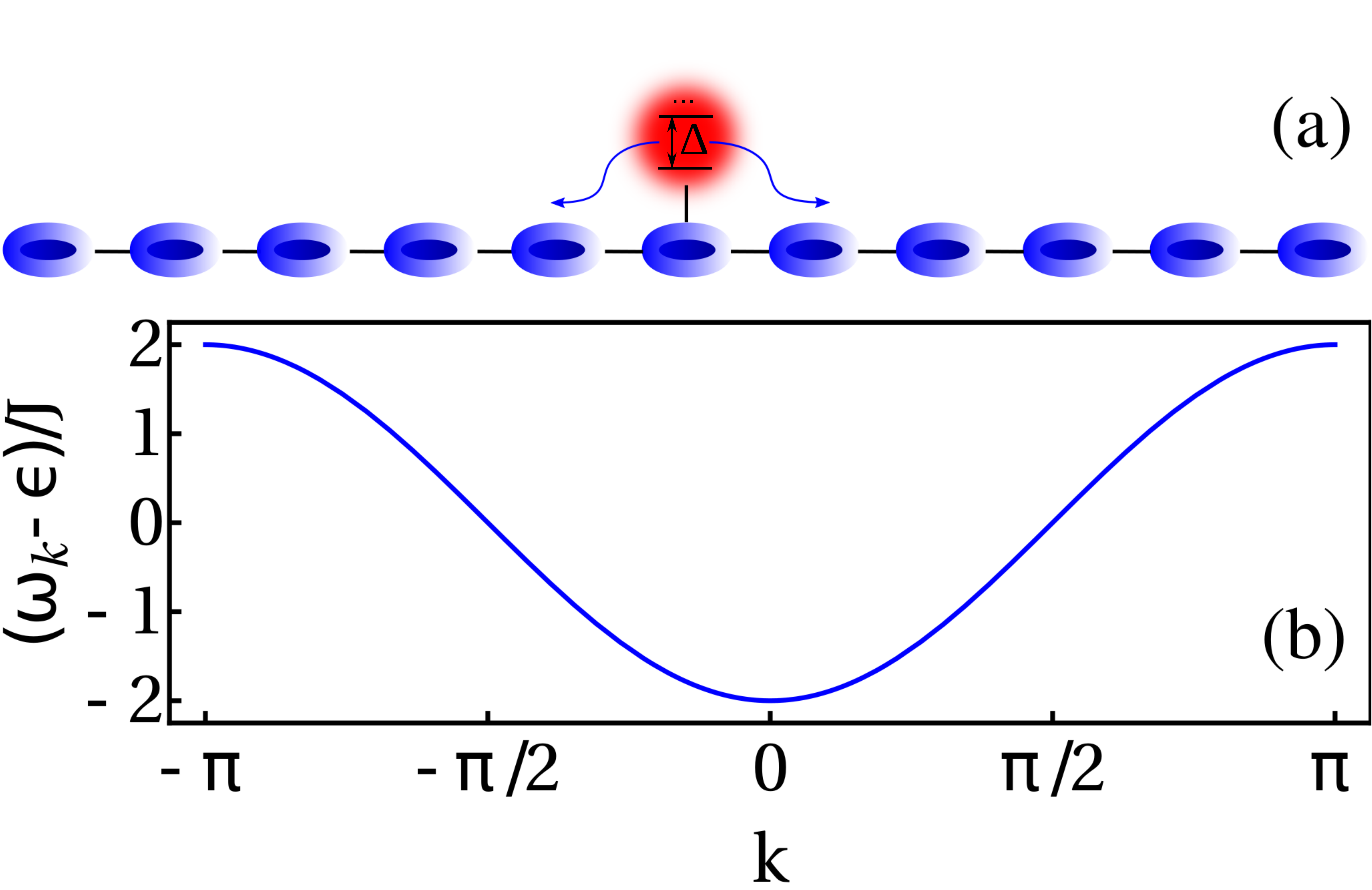}
\caption{ {\bf (a) Scheme of the system.} In blue, the bosonic coupled-cavity array. The impurity is represented as a blurred red circle. The exciton energy is $\Delta$. {\bf (b) Dispersion relation for the bosonic array}. Dispersion relation $\omega_k$ as a function of the dimensionless momentum $k$.}\label{fig:scheme}
\end{figure}

The photonic medium is an infinite chain of discrete bosonic sites coupled to an impurity placed at site $x_0=0$. 
The Hamiltonian of the combined system is ($\hbar =1$)
\begin{align}
\label{eq:H} H   = \; &  
\Delta b^\dagger b  + 
\sum_{x=-\infty}^\infty \left(\epsilon a^\dagger_x a_x -  J ( a_{x+1}^\dagger a_x  + a_{x}^\dagger a_{x+1})\right)
 \nonumber\\
&  + g ( b^\dagger \,  a_0 +   a_0^\dagger \, b) \, ,
\end{align}
where $a_{x}$ and $a_{x}^\dagger$  annihilate and create, respectively, a photon at position $x$ and $b$ and $b^\dagger$ annihilate and create excitations at the impurity. This impurity can be a two-level atom or qubit, another resonator, a spin, or any system equivalent to a qubit in the single-particle subspace.
The energy of the excited state of the impurity is $\Delta$. From now on, we borrow the nomenclature from molecular physics and refer to $\Delta$ as the exciton energy; in the same way, $b$ and $b^\dagger$ will annihilate and create an exciton.
The band of free photons is defined by a dispersion relation which depends on both the on-site photon energy $\epsilon$ and the hopping parameter $J$:  $\omega_k = \epsilon  - 2 J \cos k$, being $k$ the dimensionless momentum and $\omega_k$ the corresponding energy. In consequence, the bandwidth is $4J$. The momentum $k$ lies in $[-\pi,\pi)$. 
The group velocity is $v_k\equiv d\omega_k/dk=2J\sin k$. 
The interaction Hamiltonian (the last term in Eq.\ \eqref{eq:H}) is the dipole-field Hamiltonian in the rotating-wave approximation (RWA), which is given by the celebrated Jaynes-Cummings model, where $g$ is the coupling constant. A scheme of the system and the dispersion relation $\omega_k$ are shown in Figs. \ref{fig:scheme}(a) and (b), respectively. This model can be realized with the instances of quantum technologies enumerated in the Introduction.

Due to the rotating-wave approximation, the Hamiltonian \eqref{eq:H} commutes with the number operator $\mathcal{N}\equiv \sum_x a_x^\dagger a_x + b^\dagger b$. Thanks to this symmetry, this model is analytically solvable in the single-excitation subspace. A complete basis is formed by the scattering eigenstates $|\Psi_k\rangle$ \cite{Nori2008a} and the bound states $|\Psi_\pm\rangle$ \cite{Longo2010,Longo2011}. We introduce now the bound states and we leave the scattering ones for the next subsection.
They read
\begin{equation}
 |\Psi_\pm\rangle =  N_\pm \left(\sum_x e^{-\kappa_\pm |x|} a_x^\dagger + d_\pm b^\dagger\right)|0\rangle.\label{eq:bound_states}
\end{equation} 
The state $|0\rangle$ represents the vacuum state of the system ($a_x|0\rangle = b|0\rangle = 0$). The factor $N_\pm$ is a normalization constant, $1/|\kappa_\pm|$ is the localization length, and $d_\pm$ is  the exciton amplitude. The energy of $|\Psi_\pm\rangle$ is $\omega_\pm = \epsilon - J(e^{-\kappa_\pm} + e^{\kappa_\pm})$. The expressions of $d_\pm$ and $N_\pm$, as well as the computation of $\kappa_\pm$, are given in App. \ref{app:eigen}. The quantities $\kappa_\pm$ 
fix the properties of the bound states; namely, their energies $\omega_\pm$, exciton amplitudes $d_\pm$, and normalization factors $N_\pm$.

\begin{figure}[thb!]
\includegraphics[width=1\columnwidth]{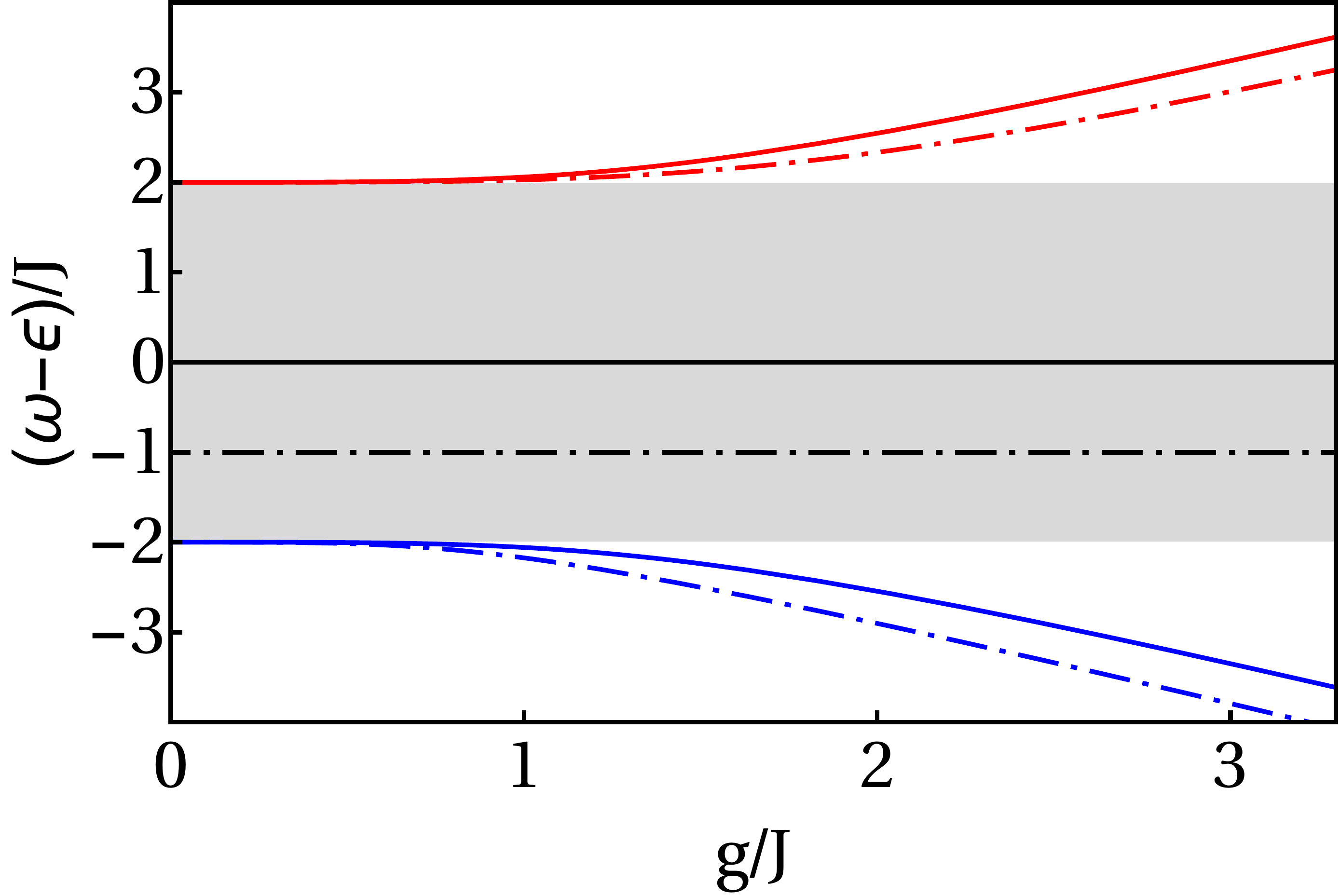}
\caption{{\bf Bound states.} Bound-state energies $(\omega_\pm - \epsilon)/J$ for two cases: $(\Delta -\epsilon)/J = 0$ (solid lines) and $(\Delta -\epsilon)/J=-1$ (dotted-dashed lines). The red upper curves are for $\omega_+$ and the blue lower ones for $\omega_-$. As a reference, the values of $(\Delta -\epsilon)/J=0$ and $(\Delta -\epsilon)/J=-1$ are represented by the solid and dotted-dashed black mid lines. The photonic band is shown by the shaded region.}\label{fig:E_bound}
\end{figure}

We plot the bound-state energies $\omega_\pm$ as a function of the coupling constant $g$, as well as the band limits in Fig. \ref{fig:E_bound}. 
Two cases are shown: (i) the exciton energy $\Delta$ at the middle of the band ($\Delta-\epsilon=0$, solid lines) and (ii) $\Delta$ closer to the band bottom  ($\Delta-\epsilon=-J$, dotted-dashed lines). 
The energies of the bound states lie outside of the band, thus they are localized (not propagating).
As $g\to 0$, $\omega_{-(+)}$   approaches the bottom (top) of the band.
If the exciton energy coincides with the band center, the energies of the bound states are  symmetrically located.
Otherwise, if the exciton energy is below the center, $\Delta-\epsilon<0$, the energy of the lower bound state $\omega_-$ moves away from the exciton energy $\Delta$ faster than the energy of the upper bound state $\omega_+$ does, and vice-versa. Therefore, the position of the exciton energy with respect to the band center originates an asymmetry between $\omega_+$ and $\omega_-$.

\subsection{One-photon scattering}\label{sec:scatt}

\begin{figure}[thb!]
\begin{center}
\includegraphics[width=1.\columnwidth]{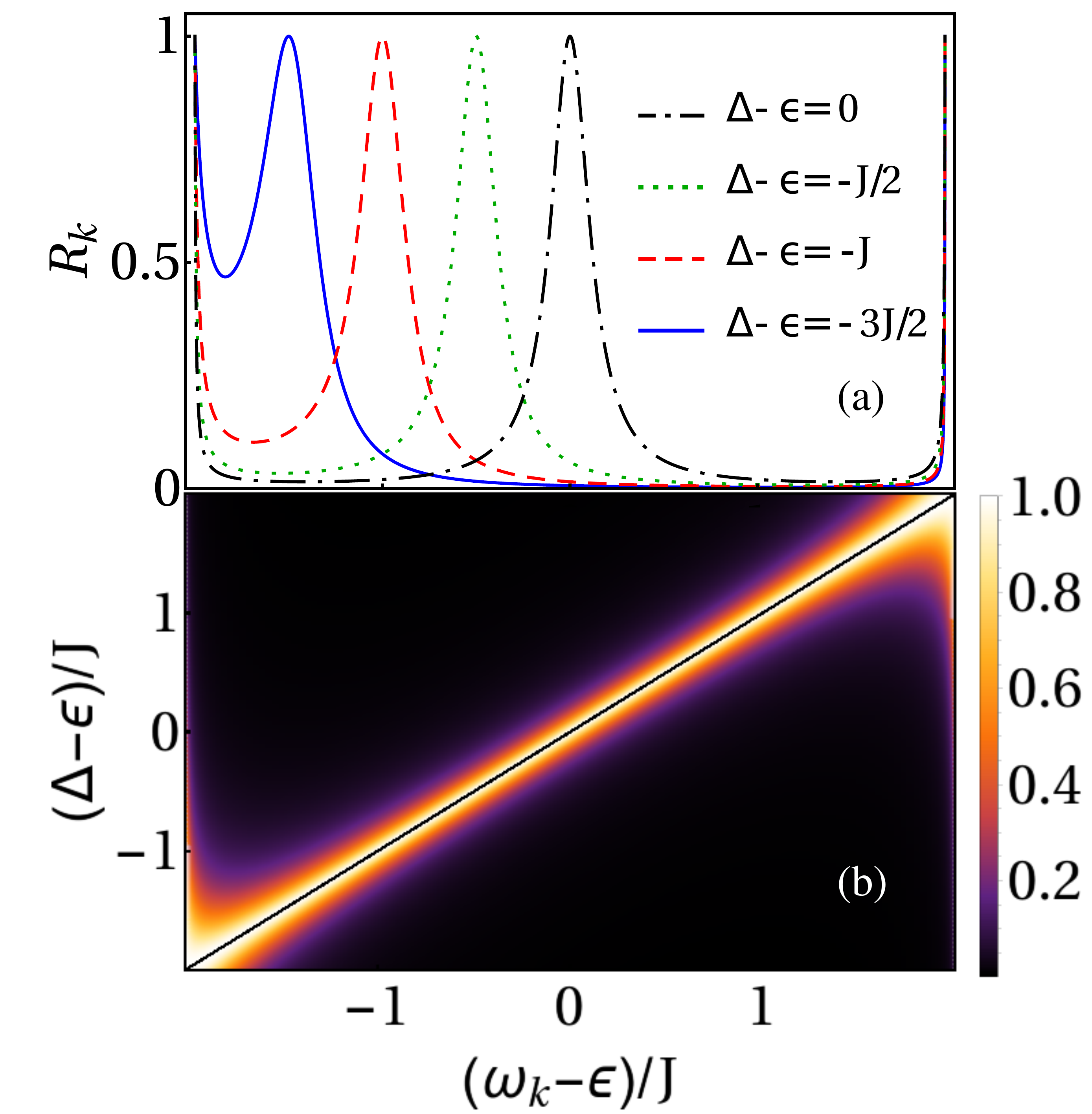}
\caption{{\bf Reflection probability.} (a) Reflection $R_k$ as a function of $(\omega_k-\epsilon)/J$ for several values of $\Delta$. (b) Reflection $R_k$ as a function of $(\omega_k-\epsilon)/J$ and $(\Delta-\epsilon)/J$ for $g=J/2$. The black line is $\Delta=\omega_k$, where $R_k=1$ (maxima in panel (a)). Notice that $R_k=1$ also at the band edges (Eq. \eqref{eq:transmission}). Notice both graphics share horizontal axis.}\label{fig:R}
\end{center}
\end{figure}

Let us now review the form of the single-particle scattering eigenstates of \eqref{eq:H} and their physical implications. They read \cite{Nori2008a}
\begin{align}
\label{eq:scattering_states} 
|\Psi_k\rangle = & \Big [ \sum_{x<0}(e^{ikx}+r_k e^{-ikx})a_x^\dagger 
 +  \sum_{x\geq 0} t_k e^{ikx} a_x^\dagger 
+ d_k b^\dagger \Big]  |0\rangle.
\end{align}
The coefficients 
 $t_k$ and $r_k$ are the  
 transmission and reflection amplitudes for an incident plane wave, respectively.  They are given by, 
\begin{align}
\label{eq:transmission}
t_k & =\frac{iv_k(\omega_k - \Delta)}{iv_k(\omega_k-\Delta)-g^2} \, , 
\\
\label{eq:reflection}
r_k&=t_k-1\, ,
\\ 
d_k  &= \frac{g t_k}{\omega_k-\Delta} \,
\label{eq:d_scattering_states}.
\end{align} 
A well-known feature in this system \cite{Fan2005a,Fan2005b,Nori2008a,Guinea1987,Roy2017} is that it presents perfect reflection, $R_k\equiv |r_k|^2 = 1$, if the energy of the input photon is equal to $\Delta$, see Eqs. \eqref{eq:transmission} and \eqref{eq:reflection}.  
This is illustrated in Figs. \ref{fig:R}(a) and (b), where $R_k$ is plotted as a function of $(\omega_k-\epsilon)/J$ for several values of $\Delta$ and as a function of $(\omega_k-\epsilon)/J$ and $(\Delta-\epsilon)/J$, respectively.
Considering the  input as a single-photon-spectroscopy probe, we could be tempted to argue that, like in scattering, the impurity emission is also maximum at resonance.  We will show that, due to the presence of bound states, this is not the case.

\section{Spontaneous decay}\label{sec:spontaneous_decay}

We discuss now the spontaneous emission of the exciton. For that, we consider that the impurity is excited at $t=0$, $|\Psi(0)\rangle = b^\dagger |0\rangle$, and compute the time evolution of the system. Spanning this state in bound and scattering eigenstates, Eqs. \eqref{eq:bound_states} and \eqref{eq:scattering_states}, respectively, the state at time $t$ is
\begin{align}
|\Psi(t)\rangle = & \int_{-\pi}^\pi \frac{dk}{2\pi}\;c_ke^{-i\omega_k t}|\Psi_k\rangle \nonumber \\
 + & c_+ e^{-i\omega_+ t} |\Psi_+\rangle + c_- e^{-i\omega_- t} |\Psi_-\rangle, \label{eq:psi(t)}
\end{align}
with
\begin{align} \label{eq:ck}
c_k &  = d_k^* =  \frac{iv_k g}{iv_k(\omega_k - \Delta) + g^2},\\
\label{eq:cpm}
c_\pm & =  \left(\frac{1+e^{-2\kappa_\pm}}{1-e^{-2\kappa_\pm}}+\frac{g^2}{(\omega_\pm - \Delta)^2}\right)^{-\frac{1}{2}} \frac{g}{\omega_\pm - \Delta}.
\end{align}

In the following, we exploit these formulae to obtain our results. We first discuss the behavior of the mean energy of the emitted wave packet. Then, we describe the spatial profile of that photon depending on $\Delta$. Finally, we study the dynamics of the exciton.

\subsection{Energy shift}

The state given by Eq. \eqref{eq:psi(t)} can be used to obtain the average value of the Hamiltonian \eqref{eq:H}. As it is a conserved quantity, it must be equal to the value at $t=0$, which is $\Delta$:
\begin{equation}\label{eq:H(t)}
\langle H\rangle = \Delta  = \int_{-\pi}^\pi \frac{dk}{2\pi} |c_k|^2 \omega_k + |c_+|^2 \omega_+ + |c_-|^2 \omega_- \, .
\end{equation}
The  average energy for the propagating field  is
\begin{equation}
\omega_\text{ph} \equiv \frac{\int_{-\pi}^\pi \omega_k |c_k|^2 dk/2\pi}{\int_{-\pi}^\pi |c_k|^2 dk/2\pi} = \frac{\int_{-\pi}^\pi \omega_k |c_k|^2 dk/2\pi}{(1-P_\text{lig})},
\end{equation}
with $P_\text{lig} \equiv |c_+|^2 + |c_-|^2$. 
Using Eq. \eqref{eq:H(t)}, $\omega_{\rm ph}$ can be written in a more convenient way
\begin{equation}
\omega_\text{ph} =\frac{\Delta - |c_+|^2 \omega_+ - |c_-|^2 \omega_-}{1-P_\text{lig}}, \label{eq:omega_ph}
\end{equation}
which shows that the energy of the emitted photon is typically different from $\Delta$ because of the presence of the bound states. In short, the amount of energy going to the propagating states must compensate that going to the bound ones so that the total energy is conserved. This is the physical origin of the energy shift.

This confirms the nonequivalence between scattering and emission spectra, since the scattering resonance always occurs when the input energy is $\Delta$, see Eq. \eqref{eq:transmission} and Fig. \ref{fig:R}.

\begin{figure}[thb!]
\includegraphics[width=1.0\columnwidth]{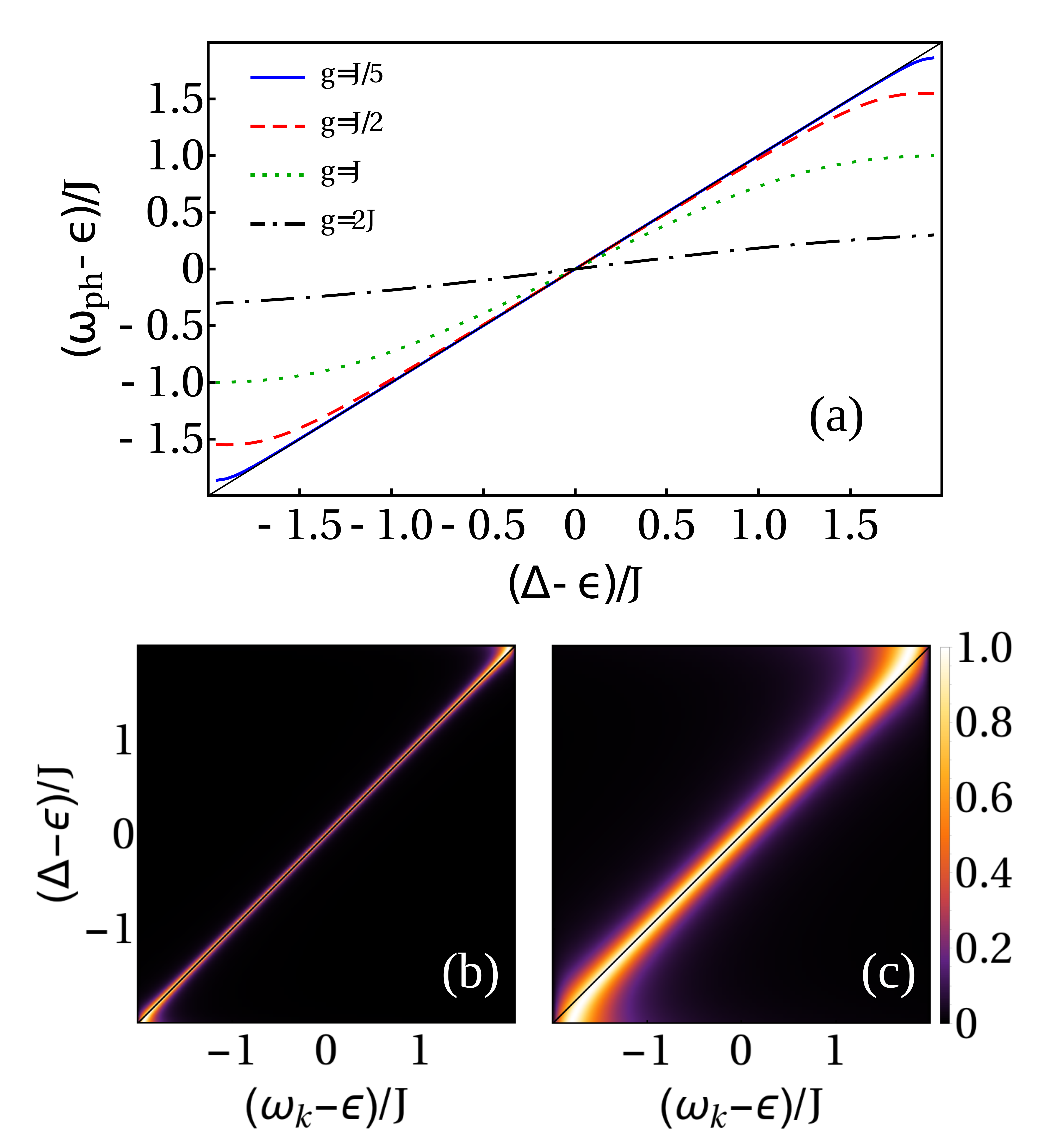}
\caption{{\bf Emitted energy.} {\bf (a)} Average energy of the emitted photon $(\omega_\text{ph}-\epsilon)/J$ as a function of $(\Delta-\epsilon)/J$ for $g=J/5,J/2,J,2J$. For reference, the straight line renders the diagonal $\omega_\text{ph}=\Delta$. In {\bf (b)} and {\bf (c)}, we plot $|c_k|^2$ as a function $(\omega_k-\epsilon)/J$ and $(\Delta-\epsilon)/J$ for $g=J/5$ and $g=J/2$, respectively. The black line renders $\Delta=\omega_k$. For each $\Delta$, we normalize $c_k$ such that $\text{max}_k(|c_k|^2)=1$.}
\label{fig:E_shift}
\end{figure}

The energy of the emitted photon $(\omega_\text{ph}-\epsilon)/J$ is plotted as a function of $(\Delta-\epsilon)/J$ in Fig. \ref{fig:E_shift}(a) for several values of $g$. The closer $\Delta$ is to the band edges, the more $\omega_\text{ph}$ departs from $\Delta$. In fact, if $\Delta$ is close to the bottom of the band, where the frequency shift is larger, the effect of the upper bound state is negligible ($|c_+|^2\ll |c_-|^2$), and vice-versa. In conclusion, the frequency shift survives in waveguides without an upper cutoff. The shift increases monotonically with $g$. Eventually, as $g/J\to\infty$, the emitted energy coincides with the middle of the band for all $\Delta$. Notice that, when the exciton energy is in the middle of the band, \emph{i.e.} when $\Delta=\epsilon$, the following relation holds: $|c_+|^2(\omega_+-\Delta)=|c_-|^2(\Delta-\omega_-)$. Inserting this in Eq. \eqref{eq:omega_ph}, we conclude that the emitted energy is equal to the exciton one, $\omega_\text{ph}=\Delta$. This is related to the symmetry of the energy of the bound states, already discussed in Sect. \ref{sec:H_BS} (see Fig. \ref{fig:E_bound}).



We also study the energy distribution of the emitted photon $|c_k|^2$. We plot it as a function of $(\omega_k-\epsilon)/J$ and $(\Delta-\epsilon)/J$ for the representative cases of $g=J/5$ and $g=J/2$ [Figs. \ref{fig:E_shift}(b) and (c), respectively]. If the coupling is small enough (left panel), the energy distribution is well peaked around $\omega_k=\Delta$. However, as $g$ increases (right panel), $|c_k|^2$ reaches its maximum for $\omega_k\neq\Delta$, being the difference larger the closer $\Delta$ is to one of the band edges. This deviation of the maximum away from $\Delta$ implies a frequency shift of the emitted photon, as already seen in Eq. \eqref{eq:omega_ph} and Fig. \ref{fig:E_shift}(a). 
The reason is simple. In the spontaneous emission some energy is released into the bound states, with a mean energy that does not generally match the exciton energy. Therefore, the coupling into flying photons must compensate for this imbalance. However, due to the fact that bound and scattering states are orthogonal, the former do not play any role in the latter. It is worthy to emphasize that this mechanism is rather general. In any photonic system supporting single-particle bound states, the frequency of the flying photon arising from spontaneous emission will present a shift with respect to that of the exciton.

\begin{figure}[thb!]
\includegraphics[width=1.0\columnwidth]{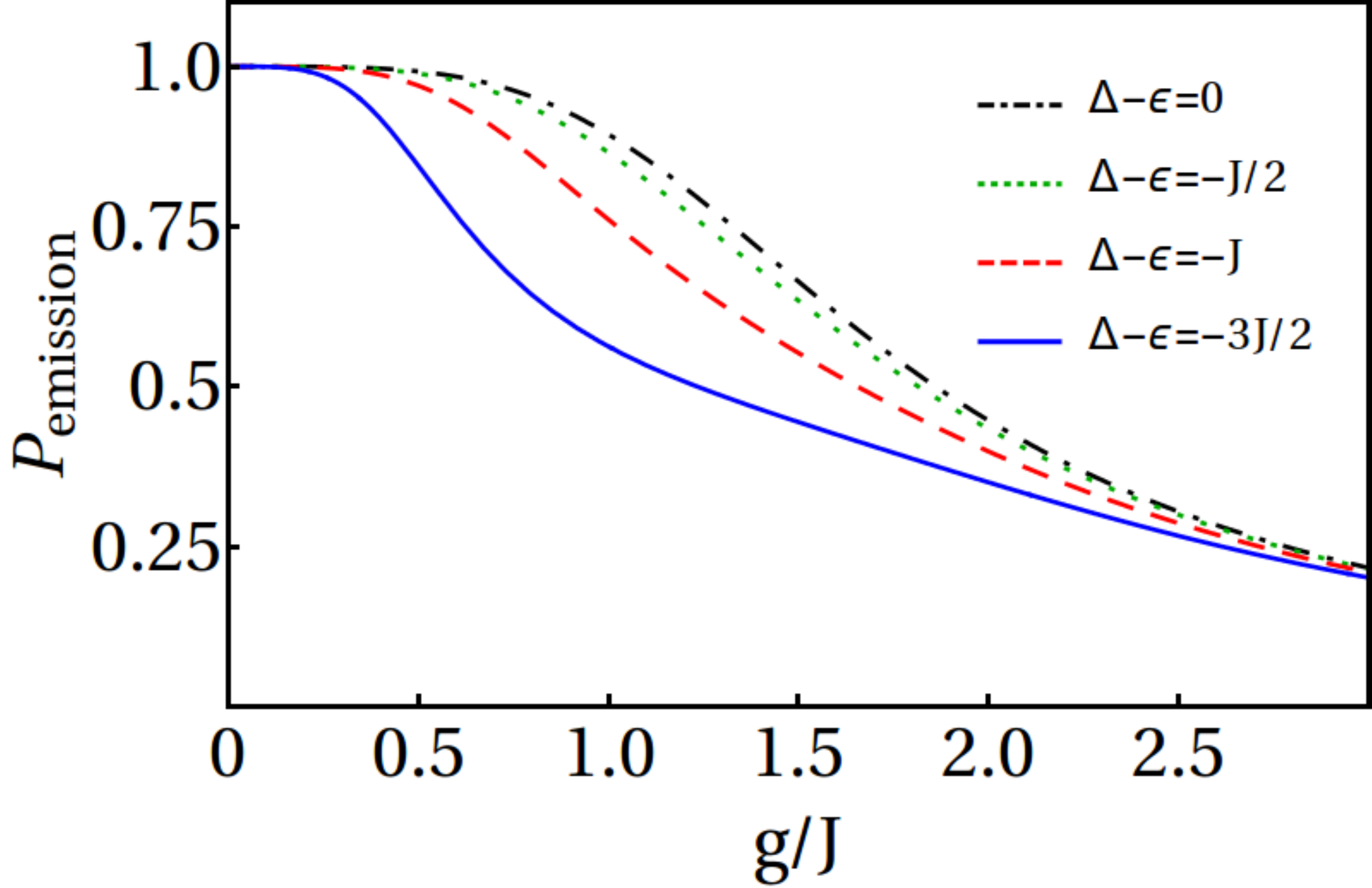}
\caption{{\bf Probability of photon emission.} Probability of emitting a flying photon, $P_{\rm emission} = 1-P_{\rm lig} = 1-|c_+|^2 - |c_-|^ 2$, as a function of $g/J$ for $\Delta-\epsilon=-3J/2,-J,-J/2,0$ from bottom to top (solid blue, dashed red, dotted green, and dotted-dashed black, respectively).}\label{fig:P_emi}
\end{figure}

We also characterize the emission probability into propagating modes, $P_\text{emission}\equiv 1-P_\text{lig} = 1 - |c_+|^2 - |c_-|^2$, in Fig. \ref{fig:P_emi}. Two effects are observed. First, the emission into bound states is negligible ($P_\text{emission} \simeq 1$) in the range $g/J \ll1$. Increasing this ratio, $P_\text{emission}$ decreases. Besides, the closer $\Delta$ is to the band gap, the smaller $P_\text{emission}$ is. Anyway, the emission probability is appreciable for really large values of the ratio $g/J$: for instance $g/J \simeq 2.5$ yields $P_\text{emission} \simeq 0.25$ for the values of $\Delta$ considered in Fig. \ref{fig:P_emi}.

\subsection{Emitted field}

\begin{figure}[thb!]
\includegraphics[width=1.0\columnwidth]{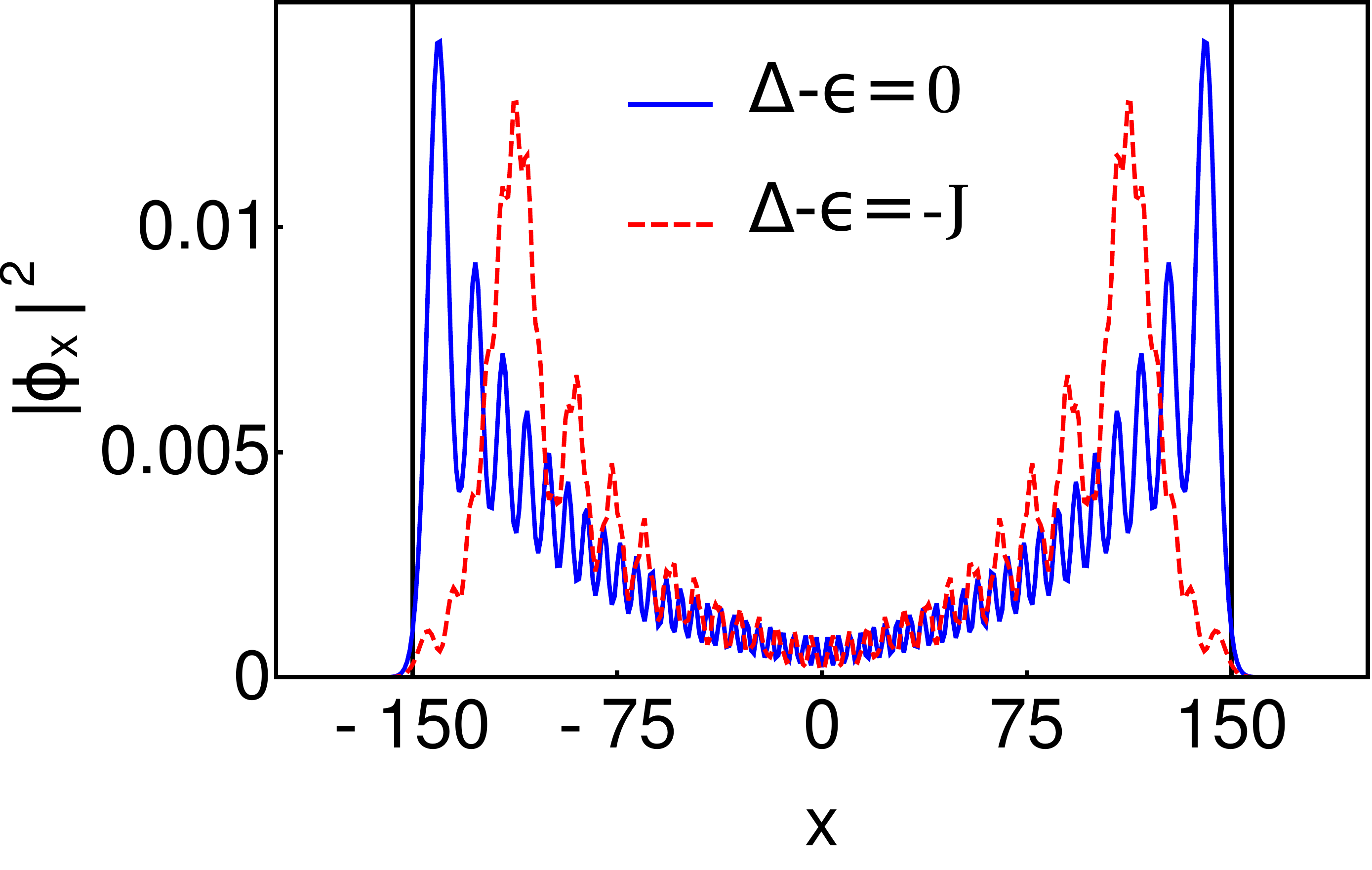}
\caption{{\bf Field distribution.} $|\phi_x|^2$ as a function of $x$ at time $t=75/J$ for $\Delta-\epsilon=0$ (solid blue) and $\Delta-\epsilon=-J$ (dashed red) for coupling $g=J/5$. The black solid vertical lines render the propagation limit $|x|=x_{\max}=v_{\max}t$, with $v_\text{max}=v_{k=\pi/2}=2J$.}\label{fig:w_n}
\end{figure}

We now study the spatial profile of the emitted field. We compute the amplitudes in position space, $\phi_x(t)\equiv \langle 0|a_x|\Psi(t)\rangle$ [Cf. App.\ \ref{app:field}]. The photon probability distribution $|\phi_x(t)|^2$ is shown in Fig. \ref{fig:w_n}, at time $t=75/J$ and $g=J/5$, for two values of the detuning: $\Delta-\epsilon=0$ (blue solid) and $\Delta-\epsilon = -J$ (red dashed). The vertical solid black lines represent $|x|=x_\text{max}\equiv v_\text{max}t$, defined in terms of the maximum group velocity $v_\text{max}= v_{k=\pi/2}=2J$. 

The probability $|\phi_x|^2$ is mostly confined within the causal cone. For $|x|>x_\text{max}$, it is not zero but it decays exponentially, as expected for the free-field scalar {\it propagator} \cite[Sect. 4.5]{Greiner-fq}, \cite[Sect. 2]{Peskin}. If $\Delta$ is in the middle of the band, the emitted photon has a momentum distribution peaked around $k=\pi/2$, where $v_k=v_\text{max}$. If $\Delta\neq\epsilon$, the velocity of the emitted photon is not peaked around $v_\text{max}$ so the maximum of $|\phi_x|^2$ is below $x_\text{max}$ (see the dashed red curve of Fig. \ref{fig:w_n}, where $\Delta-\epsilon=-J$). Lastly, notice that the emitted photon would be well peaked around $|x_\text{max}|$ in position space, independently of the value of $\Delta$, if the dispersion relation were linear.

\subsection{Impurity dynamics}

We finish with a detailed study of the exciton dynamics. From Eq. \eqref{eq:psi(t)}, we extract the time dependence of the amplitude of the exciton $b^\dagger|0\rangle$
\begin{equation}
\label{eq:qubit_amplitude}
c_{\rm e}(t) \equiv
\langle 0|b|\Psi(t)\rangle 
 =
c_{\rm e}^\text{s}(t) + 
c_{\rm e}^\text{b}(t) \, ,
\end{equation}
with $c_{\rm e}^\text{b}(t) = \sum_{\alpha=\pm}|c_\alpha|^2 e^{-i\omega_\alpha t}$ and $c_{\rm e}^\text{s}(t) = \int_{-\pi}^\pi dk |c_k|^2 e^{-i\omega_k t}/2\pi$ the contributions from the bound and scattering states repectively, see Eqs. \eqref{eq:bound_states} and \eqref{eq:scattering_states}.

First, we focus on  $c_{\rm e}^\text{s}(t)$:
\begin{equation}
c_{\rm e}^\text{s}(t) =  e^{-i\epsilon t}\frac{4g^2}{\pi J^2}\int_{-1}^1 dy\; F(y)e^{i2yJt}, \label{eq:c_sc_app}
\end{equation}
with
\begin{equation}
F(y)=\frac{\sqrt{1-y^2}}{4(1-y^2)\left((\Delta-\epsilon)/J+2y\right)^2+(g/J)^4}.\label{eq:F}
\end{equation}
\begin{figure}[thb!]
\includegraphics[width=1.0\columnwidth]{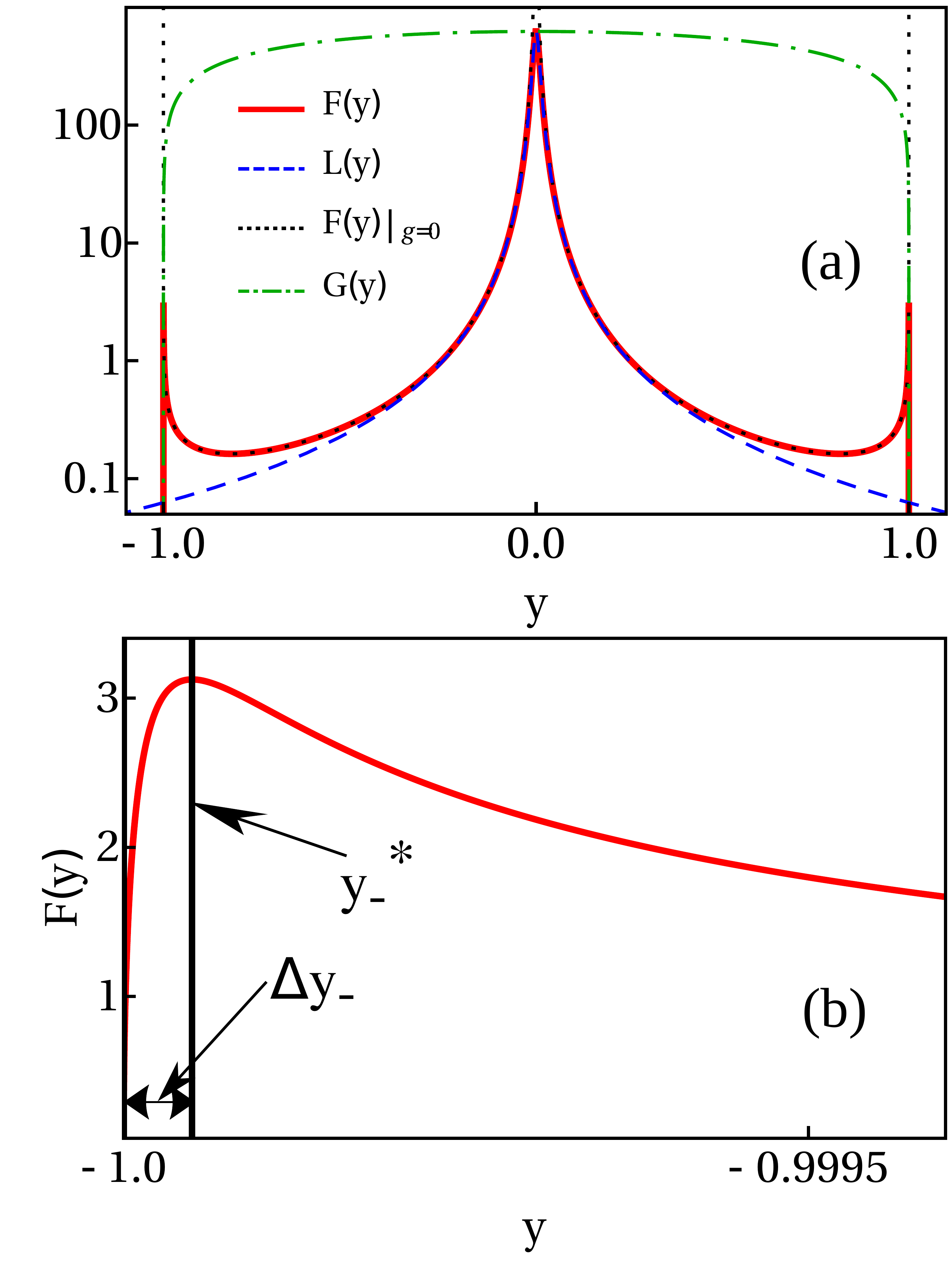}
\caption{{\bf Integrand for $c_{\rm e}^{\rm s}(t)$.} (a) Kernel $F(y)$ in logarithmic scale for $g=J/5$ (red, solid), Lorentzian approximation (blue, dashed), $F(y)$ for $g=0$ (black, dotted), and $G(y)$ (Eq. \eqref{eq:G}, black, dotted-dashed). We fix $\Delta=\epsilon$. In (b), we zoom in $F(y)$ around $y\gtrapprox -1$, with the same parameters as those used in (a). The kernel $F(y)$ reaches a maximum ay $y=y_-^*$ and $\Delta y_- = y_-^* +1$. Notice that the scale is not logarithmic in this case.}\label{fig:integrand}
\end{figure}
The behavior of $c_{\rm e}^\text{s}(t)$ is determined by the kernel $F(y)$, which is related to the density of photonic states as a function of the dimensionless energy $y = \cos k$. This kernel is plotted in Fig. \ref{fig:integrand}(a). At sufficiently long times, the oscillating term in the integral \eqref{eq:c_sc_app}, $e^{i2yJt}$, cancels out any smooth contribution of $F(y)$.
Therefore, the asymptotic relaxation dynamics is governed by the sharpest peaks and the singularities of $F(y)$. There are three main contributions: (i) a Lorentzian peak, associated to a pole of $F(y)$ in the complex plane, (ii) two peaks appearing at $y_\pm^*$, with $y_\pm^*$ close to $\pm 1$ (see Fig. \ref{fig:integrand}(b), where we zoom in $F(y)$ around $y=-1$), and (iii) the singular points at $y=\pm 1$, where the first derivative of $F(y)$ is discontinuous. All these features are clearly seen Figs. \ref{fig:integrand}(a) and (b).

\begin{figure}[thb!]
\includegraphics[width=1.0\columnwidth]{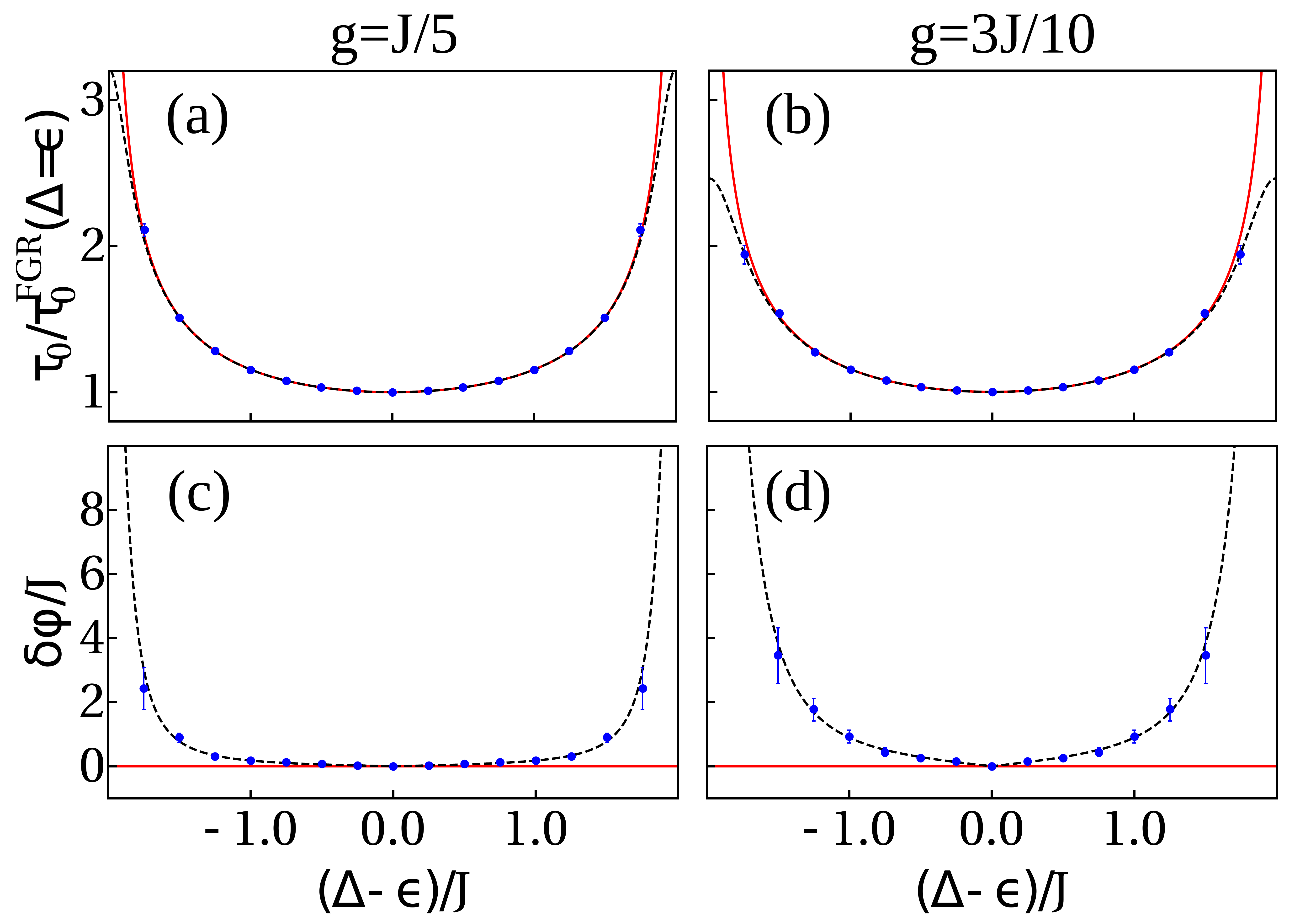}
\caption{{\bf Exponential decay.} {\bf (a), (b)} $\tau_0/\tau_0^{\rm FGR}(\Delta-\epsilon=0)$ and {\bf (c), (d)} $\delta\varphi/J$ as a function of the position of the exciton energy with respect to the band for $\epsilon=\Delta$. The coupling is $g=J/5$ (left panels) and  $g=3J/10$ (right panels). We divide $\tau_0$ by the decay time given by the Fermi's Golden Rule at the middle of the band, $\tau_0^{\rm FGR}(\Delta-\epsilon=0)$. The red solid curve and the black dashed one correspond to the Fermi's Golden Rule and to the single-pole approximation, respectively. The blue points are computed numerically; we fit the exact dynamics computed with \eqref{eq:c_sc_app} to an exponential for $t<\tau_0$.}\label{fig:qubit_decay}
\end{figure}

The Lorentzian peak gives an exponential decay $c_{\rm e}^\text{s}(t)\sim e^{-(i\varphi+1/2\tau_0)t}$. This is equivalent to an excited atom emitting photons into the free space. This contribution is the fastest and main one for short-enough times, $t<\tau_0$, since it comes from the widest peak in $F(y)$, see Fig. \ref{fig:integrand}(a).
We compare the (numerical) exact results for $\tau_0$ and $\delta\varphi\equiv\varphi-\Delta$, computed by integrating Eq. \eqref{eq:c_sc_app}, with those obtained with the Lorentzian approximation of $F(y)$ in Fig. \ref{fig:qubit_decay}. We also compare the results to those obtained with Fermi's Golden Rule: $\tau_{0}^{\rm FGR}=J\sin k_\Delta/g^2$, with $k_\Delta$ such that $\omega_{k_\Delta}=\Delta$, and $\varphi^{\rm FGR}=\Delta$. Fermi's Golden Rule describes accurately the exact results when $\Delta$ is around the middle of the band, but corrections are necessary when $\Delta$ gets closer to the band edges and when the coupling $g$ increases. The exciton energy appears in the phase of the exponential up to a correction: $\varphi = \Delta+\delta\varphi$. Thus, $\delta\varphi$ is the Lamb shift due to the coupling to the photonic bath. Notice that this Lamb shift is different from the energy shift of the emitted photon (compare Fig. \ref{fig:E_shift} to Figs. \ref{fig:qubit_decay} (c) and (d)), even though both converge to $\Delta$ in the limit $g/J\to 0$. In fact, as said, there is another characteristic energy of the system with a different behaviour: the single-photon reflection resonance, which occurs exactly at the bare excitation energy $\Delta$ (see Fig. \ref{fig:R} and Eq. \eqref{eq:reflection}). 


At later times, $t\gg \tau_0$, the singular parts of $F(y)$ are relevant. Singularities give non-exponential decays \cite{Khalfin1958,Gaveau1995}. In particular, the contribution of the peaks of $F(y)$ at $y_\pm^*$, with $y_\pm^*\simeq \pm 1$, starts to dominate. Let us define the widths of these peaks at $y_\pm^*$ as $\Delta y_\pm\equiv |y_\pm^* \mp 1|$ (see Fig. \ref{fig:integrand}(b)). For short-enough times, when $e^{i2Jyt}$ can be considered to be constant for $y\in (-1,-1+\Delta y_-)$ and $y\in (1-\Delta y_+,1)$, the kernel $F(y)$ can be approximated by setting $g=0$ (black dotted curve in Fig. \ref{fig:integrand}(a)). At $g=0$, the kernel diverges as $1/\sqrt{1-y^2}$ when $y\to\pm 1$. This kind of singularity gives an algebraic decay $t^{-1/2}$ for $c_{\rm e}^{\rm  s}(t)$. For long-enough times, when $e^{i2Jty}$ cannot be taken as a constant, we have to consider the full kernel, with the actual value of $g$. Therefore, the mentioned divergences are rounded off and the algebraic decay is modified by exponential factors. In other words, these peaks provide a contribution $c_{\rm e}^\text{s}(t) = t^{-1/2}(a_- e^{-i2Jt}e^{-t/2\tau_{1,-}} + a_+ e^{i2Jt}e^{-t/2\tau_{1,+}})$, with $\tau_{1,\pm} = (4 J \Delta y_\pm)^{-1}$. The values of the constants $a_\pm$, as well as the details on the computation, are shown in App. \ref{app:integrand}.

Eventually, these exponential contributions vanish. The only surviving contribution comes from the singularities at the band edges. There, $F(y)$ is not differentiable and gives a non-exponential (power-law) contribution for all times to  $c_{\rm e}^\text{s}(t)$, which dominates for  $t\gg \tau_0,\tau_{1,\pm}$. We show in App. \ref{app:integrand} that this contribution goes as $c_{\rm e}^\text{s}(t)\sim t^{-3/2}\cos (2Jt-3\pi/4)$. This transition between $t^{-1/2}$ and $t^{-3/2}$ decay was already discussed in \cite{Garmon2013}, but they did not see the oscillating factors, since they took the exciton energy really close to the lower part of the band, neglecting the contribution of the upper bound state. As mentioned, this decay with $t^{-3/2}$ originates from a discontinuity in the derivative of the density of photonic states and is quite common in impurity decay problems \cite{Chiu1977}, both for continuous systems \cite{Winter1961,GarciaCalderon2006} and for discrete ones \cite{Longhi2006a,Dente2008,Longhi2006b}.

The contribution of the bound states $c_{\rm e}^\text{b}(t)$ is much simpler: it gives an oscillatory term which persists for infinitely long times: $P_{\rm e}^\text{b}(t)\equiv|c_{\rm e}^\text{b}(t)|^2=|c_+|^4+|c_-|^4+2|c_+c_-|^2\cos((\omega_+-\omega_-)t)$, \cite{Gaveau1995,Lombardo2014}.

\begin{figure}[thb!]
\includegraphics[width=1.0\columnwidth]{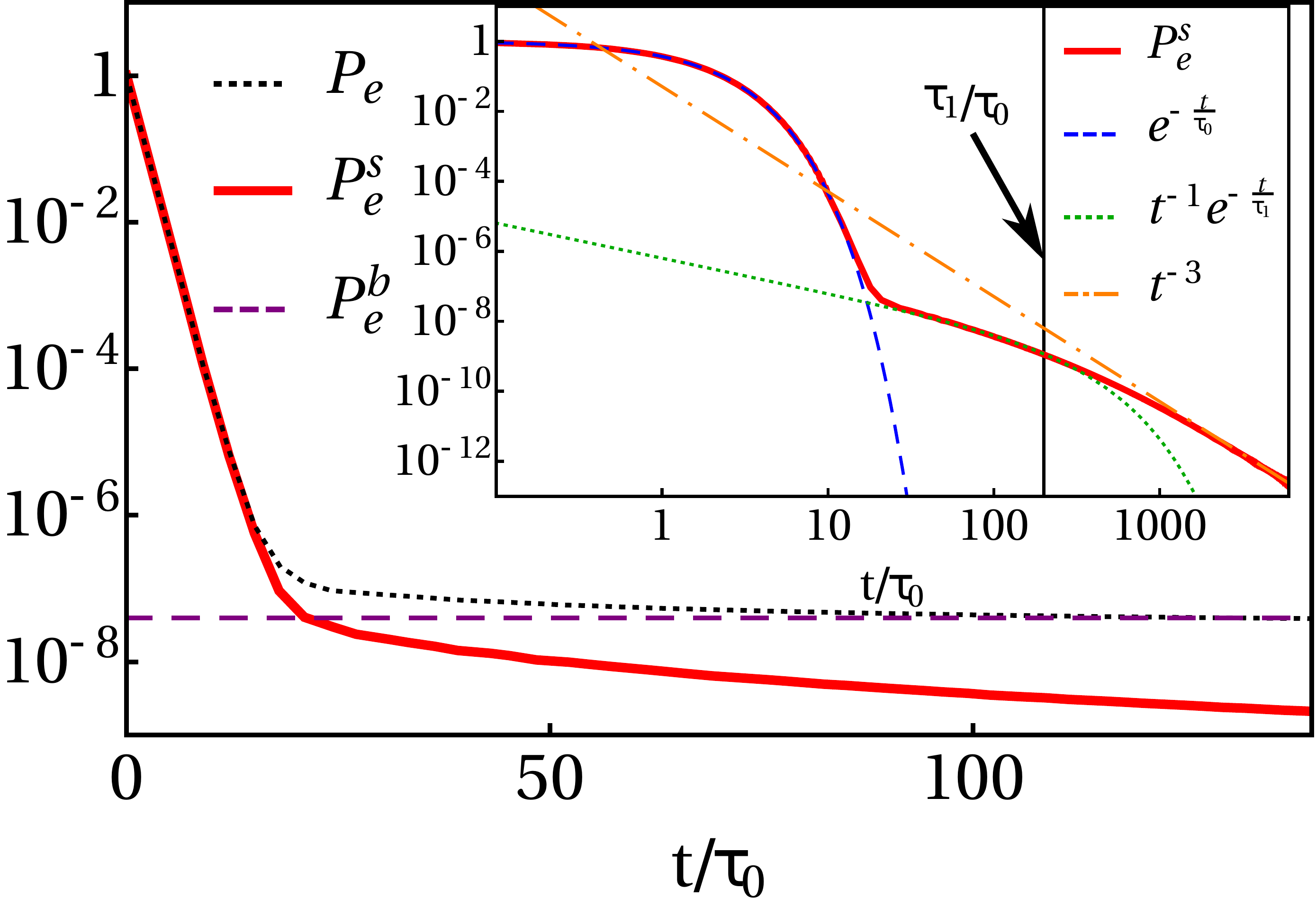}
\caption{{\bf Impurity dynamics.} $P_{\rm e}(t)$ (black, dotted), $P_{\rm e}^\text{s}(t)$ (red, solid), and $P_{\rm e}^\text{b}(t)$ (purple, dashed) for $\Delta-\epsilon=0$ and $g=J/5$ in logarithmic scale. In the inset we show $P_{\rm e}^\text{s}(t)$ in log-log scale with the three contributions: the exponential decay (blue, dashed), the power-law with $t^{-1}$ (green, dotted), and the decay with $t^{-3}$ (orange, dotted-dashed). For the sake of clarity, we average the oscillations.}\label{fig:qubit_dynamics}
\end{figure}

We sum up all this information in Fig.\ \ref{fig:qubit_dynamics}, where we plot the impurity dynamics for $\Delta-\epsilon=0$ and $g=J/5$ (same parameters as in Fig. \ref{fig:integrand}), using logarithmic scale. 
For the sake of clarity, we average the oscillations coming from the different contributions: $a_- e^{-i2Jt}+a_+ e^{i2Jt}$ (arising from the peaks around $y_\pm^*$), $\cos(2Jt-3\pi/4)$ (from the singularities at $y=\pm 1$), and $\cos((\omega_+-\omega_-)t)$ (from $c_{\rm e}^\text{b}(t)$). The population $P_{\rm e}(t)\equiv|c_{\rm e}(t)|^2$ is drawn as a black, dotted curve. It first decays as $e^{-t/\tau_0}$. In addition, the bound-state term dominates over the remaining contributions from the scattering states. Therefore, after a transient period, $P_{\rm e}(t)$ achieves the stationary regime of $P_{\rm e}^\text{b}(t)$ (purple, dashed curve; remind that we are not showing the oscillations). We also show $P_{\rm e}^\text{s}(t)\equiv |c_{\rm e}^{\rm s}(t)|^2$ in the red solid curve. After the initial exponential decay with $e^{-t/\tau_0}$, where $P_{\rm e}^\text{s}(t)\simeq P_{\rm e}(t)$, it decays sub-exponentially. To see the different contributions to this sub-exponential decay more clearly, we plot it in the inset in log-log scale. After the mentioned exponential decay with $e^{-t/\tau_0}$, it follows a decay with $t^{-1}e^{-\tau_1}$ for $\tau_0\ll t\simeq  \tau_1$ (as $\Delta-\epsilon=0$, $\tau_1\equiv\tau_{1,+}=\tau_{1,-}$; in particular $\tau_1\simeq 200 \tau_0$ for the chosen parameters). Eventually, as $t\gg\tau_1$, $P_{\rm e}^{\rm s}(t)$ goes with $t^{-3}$. The agreement between the analytical predictions (blue dashed curve for $e^{-t/\tau_0}$, green dotted curve for $t^{-1}e^{-t/\tau_1}$ and orange dotted-dashed curve for $t^{-3}$) and the exact (numerical) integration is clear in the figure.

Finally, even though we have focused on the case with $\Delta$ in the middle of the band, the mathematical analysis shown in App. \ref{app:integrand} is general, so another choice of parameters will give the same qualitative behavior.

\subsubsection{Losses}

Here we incorporate losses to the model. We add an imaginary part both to the exciton energy and the cavity energy, $\tilde{\Delta} = \Delta-i\gamma_{\rm e}/2$ and $\tilde{\epsilon} = \epsilon-i\gamma_{\rm c}/2$.

\begin{figure}[thb!]
\includegraphics[width=1.0\columnwidth]{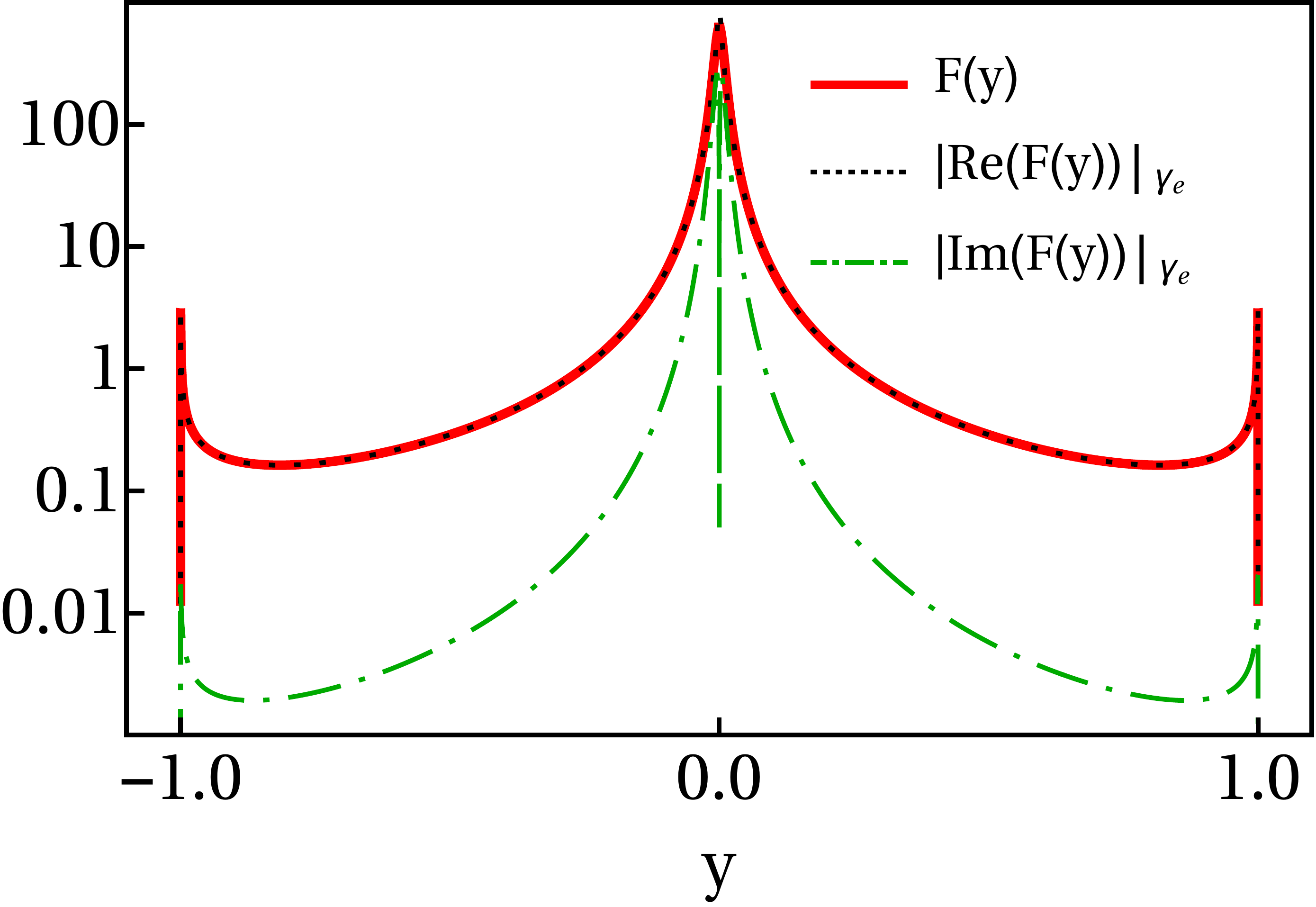}
\caption{{\bf Integrand for $c_{\rm e}^{\rm s}(t)$ with an imaginary part in $\Delta$.} Kernel $F(y)$ in logarithmic scale for $\gamma_{\rm e}=0$ (red, solid), as well as its real and imaginary part for $\gamma_{\rm e}=g/10$. The other parameters are those of Fig. \ref{fig:integrand}.}\label{fig:integrand_losses}
\end{figure}

The dynamics is still given by Eqs. \eqref{eq:c_sc_app} and \eqref{eq:F} by changing $\Delta$ and $\epsilon$ by $\tilde{\Delta}$ and $\tilde{\epsilon}$, respectively. We take $\gamma_{\rm e/c} / g \sim 0-0.15$. Considering losses in the exciton, the integrand $F(y)$ resembles to the lossless case (see Figs. \ref{fig:integrand} and \ref{fig:integrand_losses}), apart from the fact that now it is a complex function; the same happens if we instead add losses to the cavities. Therefore, we can repeat the analysis of the lossless case.

\begin{figure}[thb!]
\includegraphics[width=1.0\columnwidth]{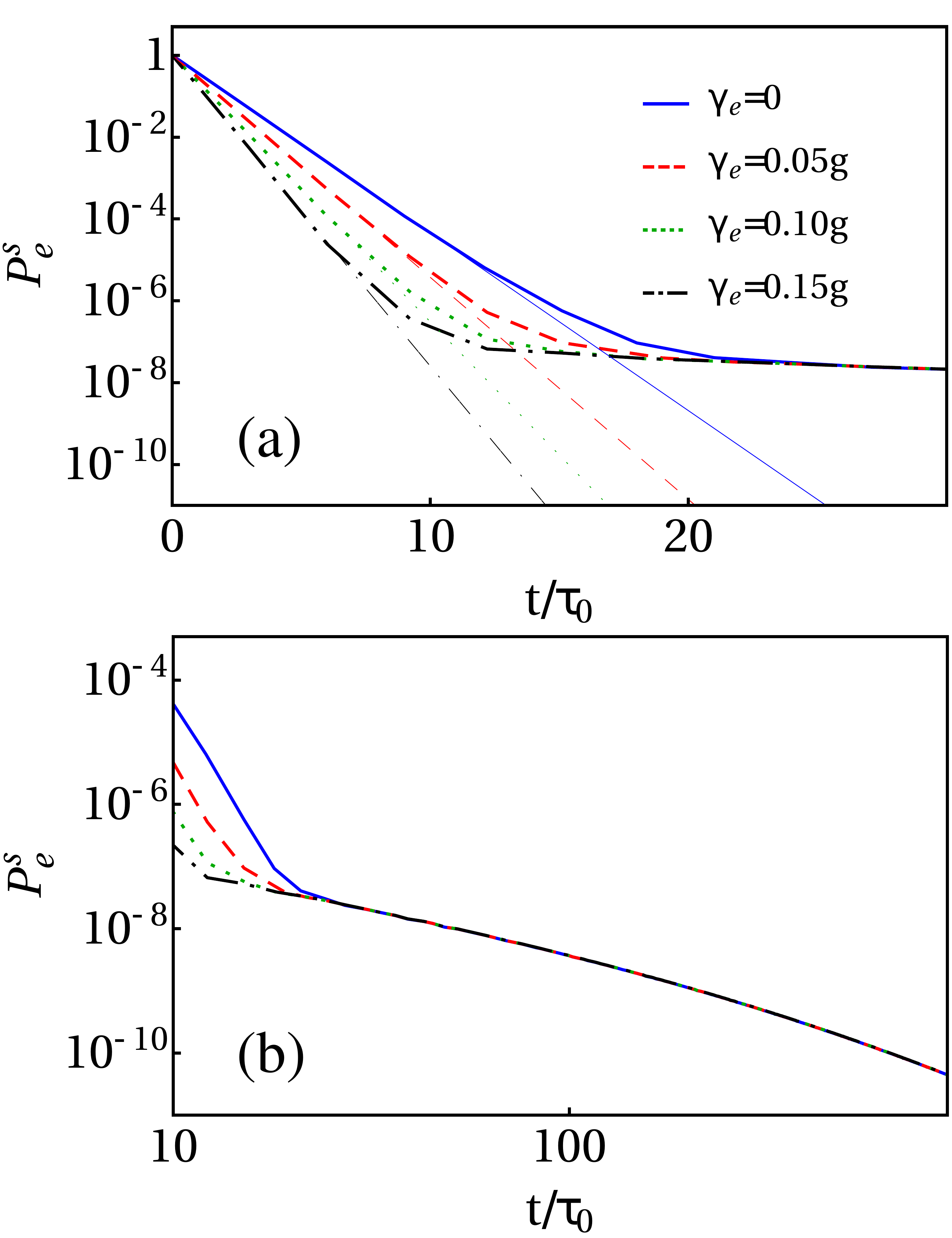}
\caption{{\bf Impurity dynamics for $\gamma_{\rm e}\neq 0$.} (a) $P_{\rm e}^{\rm s}(t)$ in logarithmic scale for several values of $\gamma_{\rm e}$. The thicker lines are the exact results, whereas the thinner ones are the analytical prediction for the exponential regime: $P_{\rm e}^{\rm s}(t)\propto e^{-(1/\tau_0 + \gamma_{\rm e})t}$. (b) The same in log-log scale and in the long-time regime. The values of $\gamma_{\rm e}$ are those of panel (a).}\label{fig:qubit_dynamics_losses}
\end{figure}

We illustrate the modifications with $\gamma_{\rm e}\neq 0$ in Fig. \ref{fig:qubit_dynamics_losses}. Initially, it still decays exponentially, but the decay rate is a sum of the previous one, $1/\tau_0$, and $\gamma_{\rm e}$: the amplitude reads $c_{\rm sc}(t) \propto e^{-(i\varphi+1/2\tau_0+\gamma_{\rm e}/2)t}$ (see Fig. \ref{fig:qubit_dynamics_losses}(a)). The power law with $t^{-1}$, $c_{\rm e}^\text{s}(t) = t^{-1/2}(a_- e^{-i2Jt}e^{-t/2\tau_{1,-}} + a_+ e^{i2Jt}e^{-t/2\tau_{1,+}})$, is preserved. The coefficients $a_\pm$, whose expressions are shown in App. \ref{app:integrand}, get modified $10^{-5}\%$ at most for the chosen values of $\gamma_{\rm e}$. Lastly, the asymptotic decay with $t^{-3}$ does not depend on $\Delta$ (see App. \ref{app:integrand}). The robustness of the power-law tails is seen in Fig. \ref{fig:qubit_dynamics_losses}(b).

\begin{figure}[thb!]
\includegraphics[width=1.0\columnwidth]{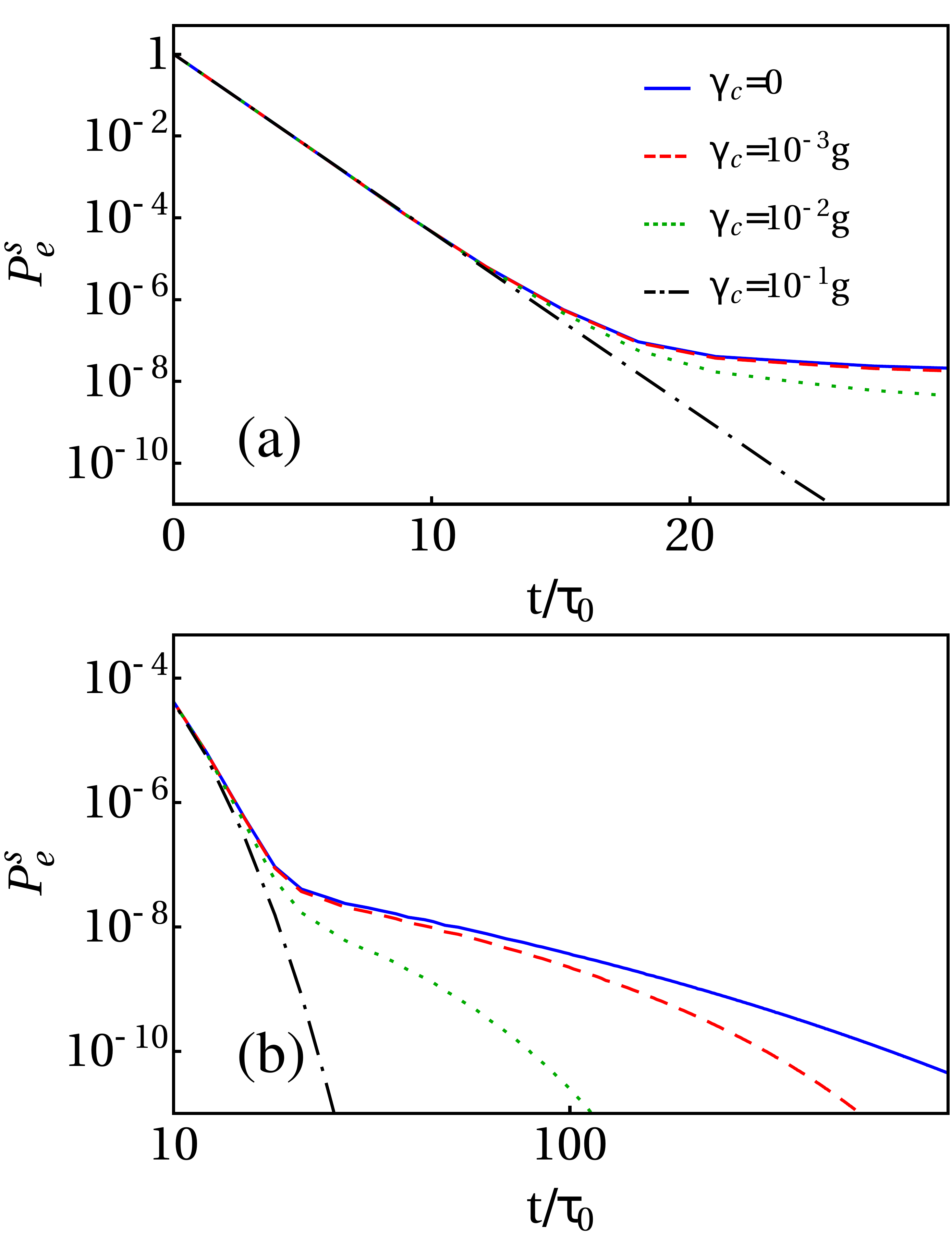}
\caption{{\bf Impurity dynamics for $\gamma_{\rm c}\neq 0$.} (a) $P_{\rm e}^{\rm s}(t)$ in logarithmic scale for several values of $\gamma_{\rm c}$. (b) The same in log-log scale and in the long-time regime. The values of $\gamma_{\rm c}$ are those of panel (a). The power laws survive for intermediate times for moderate values of $\gamma_{\rm c}$, but they disappear if $\gamma_{\rm c}$ is too large (black curve).}\label{fig:qubit_dynamics_losses_c}
\end{figure}

If we instead consider lossy cavities, $\gamma_{\rm c}\neq 0$, there is a global factor $e^{-\gamma_{\rm c}t/2}$ multiplying $c_{\rm sc}(t)$ (see Eq. \eqref{eq:c_sc_app}). When integrating $F(y)$, the imaginary part in $\epsilon$ adds an increasing exponential $e^{\gamma_{\rm c}t/2}$ to $c_{\rm sc}(t)$, contrarily to $\Delta$ (see the denominator of $F(y)$, Eq. \eqref{eq:F}; $\Delta$ and $\epsilon$ have opposite signs). This increasing exponential cancels out with the global factor $e^{-\gamma_{\rm c}t/2}$. Therefore, no modifications are seen in the initial exponential regime (see Fig. \ref{fig:qubit_dynamics_losses_c}(a)). The global factor  $e^{-\gamma_{\rm c}t/2}$ suppresses the power laws in the long-time limit. If the characteristic time of the losses $1/\gamma_{\rm c}$ is larger than $\tau_{1,\pm}$, we can see the power-law tails for intermediate times (see Fig. \ref{fig:qubit_dynamics_losses_c}(b)).

\section{Conclusions}\label{sec:conclusions}

We have discussed the differences between spontaneous-decay and scattering spectra. As we argued in the text, naively we could expect that the scattering resonance should coincide with the spontaneous-emission energy. However, whereas the scattering resonance is always equal to the exciton energy, we have shown that the emission frequency is shifted. In particular, this shift is more clear as the coupling increases and/or the exciton energy is closer to the band edges. We have also seen that the profile of the emitted photon strongly depends on the exciton energy with respect to the photonic band. Lastly, the presence of bound states and a nontrivial density of states makes the impurity dynamics nontrivial, with three dynamical regimes: exponential decay, power-law with a transition from $t^{-1}$ to $t^{-3}$, and oscillatory asymptotic regime. This dynamics has proven to be robust under the presence of losses, both in the atom and in the cavities. Even though the population at the power-law regime is very small, it could be measured. In fact, such power laws have already been measured in a context of dissolved organic materials, where the fluorescence follows an algebraic decay at long times \cite{Rothe2006}.

Some features, such as the spectroscopic shifts in the spontaneously emitted photons, can be detectable by tuning up and down the frequency of the exciton with respect to the band edge. For probing the dynamics, we suggest using a more sophisticated protocol that (i) places the exciton energy at the right frequency, (ii) then excites it and after a finite time $t$ (iii) detunes the exciton and probes dispersively its excited state population. All these ideas can be implemented in state-of-the-art setups with superconducting cavities and transmon qubits \cite{Liu2017} and also with quantum dots in photonic crystals \cite{Arcari2014,Sollner2015,Lodahl2015}.  

\begin{acknowledgements}
We acknowledge 
support by the Spanish Ministerio de Economia y Competitividad within projects MAT2014-53432-C5-1-R, FIS2015-70856-P (Cofunded by FEDER), and No. FIS2014-55867-P, the Gobierno
de Aragon (FENOL group), CAM Research Network QUITEMAD+,
and the European project PROMISCE.
\end{acknowledgements}

\appendix

\section{Bound States}\label{app:eigen}

We provide the explicit expressions for $d_\pm$, $\kappa_\pm$, and $N_\pm$ appearing in the main text (Eq. \eqref{eq:bound_states}). The excited-state amplitude of the impurity $d_\pm$ is
\begin{equation}
d_\pm = \frac{g}{\omega_\pm - \Delta}.\label{eq:d_bound_states}
\end{equation}
In order to compute $\kappa_\pm$, we define $\eta_\pm \equiv e^{-\kappa_\pm}$ and use the eigenvalue equation $H|\Psi_\pm\rangle = \omega_\pm|\Psi_\pm\rangle$ \cite{Longo2011}
\begin{equation}\label{eq:eta}
\eta_\pm ^4 + \frac{\Delta-\epsilon}{J} \eta_\pm ^3 + \frac{g^2}{J^2} \eta_\pm ^2 - \frac{\Delta-\epsilon}{J} \eta_\pm - 1 = 0.
\end{equation}
This equation has four solutions. However, we have two constrains: (i) $\text{Re}(\kappa_\pm)>0$, because the photonic cloud must be localized around the impurity and cannot explode at $x\to\pm\infty$, and (ii) $\text{Im}(\kappa_\pm)=0,\pi$, since the energies $\omega_\pm = \epsilon - J(e^{-\kappa_\pm} + e^{\kappa_\pm})$ are real. With these restrictions, there are only two solutions for $\eta_\pm$, which can be found numerically. 

If we take the limit $J\to\infty$, where the dispersion tends to be linear, the valid solutions for $\eta_\pm$ are $\pm 1$, so ${\rm Re}(\kappa_\pm) = 0$. Therefore, $\ket{\Psi_\pm}$ are not bound anymore. In fact, they converge to the scattering states $\ket{\Psi_k}$ with $k=0$ and $k=\pi$, that is, those at the band edges.

The normalization factor is
\begin{equation}\label{eq:Npm}
N_\pm = \left(\frac{1+e^{-2\kappa_\pm}}{1-e^{-2\kappa_\pm}}+|d_\pm|^2\right)^{-1/2}.
\end{equation}
Finally, $c_\pm=\langle 0|\sigma^-|\Psi_\pm\rangle=(N_\pm d_\pm)^*$ can be obtained (Eq. \eqref{eq:cpm}), since we know both $d_\pm$, Eq. \eqref{eq:d_bound_states}, and $N_\pm$, Eq. \eqref{eq:Npm}.

\section{Emitted field}\label{app:field}

The profile of the emitted field $\phi_x(t)=\langle 0|a_x|\Psi(t)\rangle$ is given by
\begin{align}
\phi_x(t)& = \frac{1}{2\pi}\int_{-\pi}^\pi dk c_k e^{-i\omega_k t}\langle 0|a_x|\Psi_k\rangle \\
& + c_+ e^{-i\omega_+ t} \langle 0|a_x|\Psi_+\rangle+ c_- e^{-i\omega_- t} \langle 0|a_x|\Psi_-\rangle, \nonumber
\end{align}
where we have used Eq. \eqref{eq:psi(t)}. In order to compute the amplitude $\langle 0|a_x|\Psi_k\rangle$ we take the expression of $|\Psi_k\rangle$, Eq. \eqref{eq:scattering_states}, for $k>0$:
\begin{equation}
\langle 0|a_x|\Psi_k\rangle = \left\{ 
\begin{array}{c}
e^{ikx}+r_ke^{-ikx}\quad x<0,\\
t_k e^{ikx} \qquad\qquad\;\;\; x\geq 0.
\end{array}
\right.
\end{equation}
If $k<0$:
\begin{equation}
\langle 0|a_x|\Psi_k\rangle = \left\{ 
\begin{array}{c}
t_k e^{ikx} \qquad\qquad\;\;\; x< 0,\\
e^{ikx}+r_ke^{-ikx}\quad x\geq 0.
\end{array}
\right.
\end{equation}

The amplitudes $\langle 0|a_x|\Psi_\pm\rangle$ are computed by projecting on $|\Psi_\pm\rangle$ (Eq. \eqref{eq:bound_states}):
\begin{equation}
\langle 0|a_x|\Psi_\pm \rangle= N_\pm e^{-\kappa_\pm |x|}.
\end{equation}

\section{Impurity dynamics: analyzing the integrand}\label{app:integrand}

\subsection{Exponential decay}

In order to extract the first exponential decay, we can approximate $F(y)$ by $L(y) =  a_p/(y-y_p)$, being $y_p$ the pole corresponding to the peak of $F(y)$, with $-1<\text{Re}(y_p)<1$ and $\text{Im}(y_p)>0$, and $a_p$ the residue of $F(y)$ at $y=y_p$. The value of $y_p$ is found numerically, equating the denominator of $F(y)$ to 0 (see Eq. \eqref{eq:F}). The residue $a_p$ is computed by definition. We extend the integration domain to $\pm\infty$. Then, applying the residue theorem
\begin{equation}
c_{\rm e}^\text{s}(t) = i8 a_p (g/J)^2 e^{-i\epsilon t}e^{i2y_p Jt},
\end{equation}
By computing this numerically, we obtain the decay rate  $\tau_0 = (4J\;\text{Im}(y_p))^{-1}$ and the phase $\varphi = \epsilon - 2J\;\text{Re}(y_p)$, as shown in Fig. \ref{fig:qubit_decay} in the main text.


\subsection{Sub-exponential regime: $t^{-1/2}$}

The kernel $F(y)$ has a sharp behavior around $y_\pm^*$. In fact, it diverges when $y\to\pm 1$ if $g=0$. In order to take into account this contribution, we can approximate $F(y)$ by $F(y)|_{g=0}$ (see blue, dashed curve of Fig. \ref{fig:integrand}(a))
\begin{equation}
c_{\rm e}^\text{s}(t)\simeq \frac{4g^2 e^{-i\epsilon t}}{\pi J^2}\int_{-1}^1 dy \frac{e^{i2yJt}}{4\sqrt{1-y^2}((\Delta-\epsilon)/J+2y)^2}.
\end{equation}
If $2\Delta y_\pm Jt\ll 1$, with $\Delta y_\pm=|y_\pm^*\mp 1|$, the oscillatory term $e^{i2yJt}$ will not be sensitive to the difference between $F(y)$ and $F(y)|_{g=0}$ when $y$ is close to the edges. As we are concerned in the contribution around $\pm 1$, we can approximate the integral as:
\begin{align}
c_{\rm e}^\text{s}(t)  \simeq  \frac{4g^2 e^{-i\epsilon t}}{\sqrt{2}\pi J^2}&\left(\frac{J^2}{(\Delta-\epsilon-2J)^2}\int_{-1}^\infty dy \frac{e^{i2yJt}}{4\sqrt{1+y}}\right. \\
& \left.+ \frac{J^2}{(\Delta-\epsilon+2J)^2}\int_{-\infty}^1 dy \frac{e^{i2yJt}}{4\sqrt{1-y}}\right).\nonumber
\end{align}
These integrals are analytical
\begin{equation}
c_{\rm e}^\text{s}(t)\simeq \frac{g^2e^{-i\epsilon t}}{2\sqrt{2\pi Jt}}\left(\frac{e^{-i2Jt}}{(\Delta-\epsilon-2J)^2}+\frac{e^{i2Jt}}{(\Delta-\epsilon+2J)^2}\right).\label{eq:csc_1}
\end{equation}
In consequence, $P_{\rm e}^\text{s}(t)$ decays with $(Jt)^{-1}$ after the initial exponential decay if $\tau_0\ll t\ll \tau_{1,\pm}$, with $\tau_{1,\pm} = (4J \Delta y_\pm )^{-1}$. We can rewrite the last expression by adding the decaying exponentials with $\tau_{1,\pm}$:
\begin{align}
c_{\rm e}^\text{s}(t)\simeq \frac{g^2e^{-i\epsilon t}}{2\sqrt{2\pi Jt}}&\left(\frac{e^{-i2Jt}}{(\Delta-\epsilon-2J)^2} e^{-t/2\tau_{1,-}}\right. \nonumber \\
&\left. +\frac{e^{i2Jt}}{(\Delta-\epsilon+2J)^2} e^{-t/2\tau_{1,+}} \right).
\end{align}
The constants $a_\pm$ introduced in the main text can be identified as
\begin{align}
\label{eq:am}&a_-=\frac{g^2}{2\sqrt{2\pi J}(\Delta-\epsilon-2J)^2},\\
\label{eq:ap}&a_+=\frac{g^2}{2\sqrt{2\pi J}(\Delta-\epsilon+2J)^2}.
\end{align}

\subsection{Sub-exponential regime: $t^{-3/2}$}

Eventually, when $t\gg \tau_0, \tau_{1,\pm}$, the only surviving contribution will come from the singularities of $F(y)$, since the rapidly oscillating term $e^{i2yJt}$ cancels out the contribution of any non-singular part of the kernel. The singularities of $F(y)$ occurs at $y=\pm 1$. Therefore, we can approximate the kernel by any function which behaves as $F(y)$ for $y=\pm 1$, provided this function has no singularities in between both points. We consider the function $G(y)$ (see Fig. \ref{fig:integrand}(a), green, dotted-dashed curve)
\begin{equation}
G(y)=\frac{\sqrt{1-y^2}}{(g/J)^4}.\label{eq:G}
\end{equation}
Integrating this, the scattering amplitude in the long-time limit, $t\gg \tau_0,\tau_{1,\pm}$, is
\begin{equation}
c_{\rm e}^\text{s}(t) \simeq  \frac{4J^2e^{-i\epsilon t}}{\pi g^2}\int_{-1}^1 dy \sqrt{1-y^2} e^{i2yJt}=\frac{2Je^{-i\epsilon t}}{g^2 }\frac{J_1(2Jt)}{t},
\label{eq:c_sc_bessel}
\end{equation}
being $J_1$ the first-kind Bessel function with $n=1$. As $t\to\infty$, $J_1(2Jt)\to (\pi Jt)^{-1/2}\cos(2Jt-3\pi /4)$, so $P_{\rm e}^\text{s}(t)$ decays with $(Jt)^{-3}$ in the long-time limit.

\bibliographystyle{apsrev4-1}
\bibliography{scattering_eduardo} 

\end{document}